\documentclass[%
reprint,
 amsmath,amssymb,
 aps,
]{revtex4-2}
\usepackage{CJK}

\usepackage{graphicx}
\usepackage{caption}
\usepackage{dcolumn}
\usepackage{comment}
\usepackage{xcolor}
\usepackage[colorinlistoftodos]{todonotes}
\usepackage[colorlinks=true,citecolor=blue,urlcolor=blue]{hyperref}
\usepackage[english]{babel}
\usepackage{bm}
\usepackage{amsmath}
\usepackage{amssymb}
\usepackage{comment}
\usepackage{subfigure}
\usepackage{epstopdf}
\usepackage{appendix}
\usepackage{float}

\usepackage{mathtools}
\DeclareUnicodeCharacter{03B2}{\ensuremath{\beta}}
\UseRawInputEncoding

\begin{document}

\title{Adiabatic and post-adiabatic hyperspherical treatment of the huge ungerade proton-hydrogen scattering length}

\author{Shayamal Singh}
 \email{singh905@purdue.edu}
 \affiliation{
 Department of Physics and Astronomy, Purdue University, West Lafayette, Indiana 47907 USA
}

\author{Chris H. Greene}%
 \email{chgreene@purdue.edu}
\affiliation{%
 Department of Physics and Astronomy, Purdue University, West Lafayette, Indiana 47907 USA
}%
\affiliation{Purdue Quantum Science and Engineering Institute, Purdue University, West Lafayette, Indiana 47907 USA}

\begin{abstract}
While the hydrogen molecular ion is the simplest molecule in nature and very well studied in all of its properties, it remains an interesting system to use for explorations of fundamental questions.  One such question treated in this study relates to finding an optimal adiabatic representation of the physics, i.e. the best adiabatic description that minimizes the role of nonadiabatic effects.  As a test case explored here in detail, we consider the ungerade symmetry of H$_2^+$, which is known to have a huge scattering length of order $750$ bohr radii, and an incredibly weakly bound excited state.  We show that a hyperspherical adiabatic description does an excellent job of capturing the main physics.  Our calculation yields a competitive scattering length and shows that nonadiabatic corrections are small and can even be adequately captured using the post-adiabatic theory of Klar and Fano.
\end{abstract}

\maketitle

\section{\label{sec:level1}Introduction}
H$_2^+$ molecular ion has been a subject of extensive research due to its fundamental importance in atom scattering \cite{bates1952electron,hunter1966born} and understanding molecular bonding \cite{bates1953k,bates1983theory,lepp2002atomic,stancil1993radiative}. While the Born-Oppenheimer approximation provides a simplified system description, it remains approximate. Two electronic energy curves are associated with the first dissociation limit: the $1s\sigma_g$ curve, featuring a deep energy well that supports twenty $L = 0$ vibrational states (where $L$ represents orbital angular momentum), and the predominantly repulsive $2p\sigma_u$ curve. However, the $2p\sigma_u$ curve at large distances exhibits a weakly attractive potential. It was first shown in Ref.\cite{lazauskas2002scattering,carbonell2003new} that this potential has an excited state $(\nu=1, L=0)$. This state is extremely weakly bound, with a binding energy of $1.085045\times10^{-9}$ a.u. relative to the dissociation threshold\cite{carbonell2003new}. This manifests as an extremely large s-wave scattering length for the H$^+$ + H(1s) collision, reported to be 750(5) $a_0$ \cite{lazauskas2002scattering,carbonell2003new} neglecting hyperfine and relativistic effects .\\
The diagonal part of the hyperfine interaction was included in a later calculation along with relativistic and radiative corrections \cite{carbonell2004relativistic}. The large value of scattering length was also confirmed in later calculations \cite{li2015proton,bodo2008ultra,zhou2015hyperspherical,glassgold2005h++,kaiser2013quantization}.  The rigorous treatment of the hyperfine interaction by Merkt and Beyer \cite{beyer2018hyperfine} showed that the G = 1/2 hyperfine component of the A$^+$ $(\nu=1, N=0)$ state is nonexistent. The hyperfine interaction didn't significantly impact the binding energy of the G=3/2 hyperfine component in the A$^+$ $(\nu=1, N=0)$ state, and the corresponding scattering length for the H$^+$ + H(1s, F=1) collision was calculated to be 757(7) a$_0$. 
However, the theoretical calculations yielded very similar values for the scattering length, regardless of whether the hyperfine interaction was included \cite{carbonell2004relativistic,li2015proton} or neglected \cite{bodo2008ultra,zhou2015hyperspherical,glassgold2005h++,kaiser2013quantization}. This allows us to explore this problem while neglecting relativistic and hyperfine effects.
In this paper, we study the extent to which the physics of this ungerade state very close to the dissociation threshold can be described using an adiabatic representation. The adiabatic approximation involves the separation of one coordinate $R$, which we call the adiabatic parameter, from all the other coordinates $\Omega$. The exact wave function is then written as a superposition of channel functions $\phi_{\mu}(R;\Omega)$, which are eigenfunctions of the Hamiltonian at a fixed R-value. The suitability of an adiabatic approximation is determined by how well the physical phenomenon can be described by a slow variation of these channel wave functions with respect to the chosen adiabatic parameter $R$. A good choice not only reduces the computational effort involved in the description but also sheds light on the actual physics of the problem. This can be very important in problems that are very sensitive to the actual potential curves, where a bad choice of the adiabatic parameter would require careful consideration of the non-adiabatic corrections, leading to more complex and longer computations. In more physical terms, having the most realistic description at the single potential curve level makes it easier to identify phenomena such as shape resonances or near-threshold bound states and threshold laws.  Our main focus is on using a representation which requires minimum consideration of non-adiabatic effects. The Born-Oppenheimer approximation is a candidate, but the adiabatic curves, in this case, go to the wrong dissociation threshold, which stems from the problem of electron translation factors \cite{bates1958electron,bates1962charge}. This becomes an issue when describing physics very close to the threshold. Therefore, the adiabatic hyperspherical representation is considered for this problem since the potential curves exhibit the correct dissociation limits \cite{hyperspherical_ETF,macek1987bypassing}, which is important due to the extremely weakly bound nature of the first excited vibrational state. We also apply the post-adiabatic theory of Klar and Fano \cite{klar1976post,klar1977adiabatic} to test its ability to incorporate the main non-adiabatic effects approximately into an effective potential energy curve.\\

\section{\label{sec:level2}Mathematical formulation}
\subsection{\label{sec:level}Adiabatic Hamiltonian}
This section presents the adiabatic formulation for our system. The treatment also utilizes ideas from generalized quantum defect theory \cite{GreeneRauFano1982,Rau1988pra, WatanabeGreene1980pra, BurkeQDT1998prl, Idziaszek2011njp} to compute the proton-hydrogen scattering length for the ungerade symmetry. The centre of mass degrees of freedom can be separated, and all spin-orbit and hyperfine couplings are neglected. The non-relativistic Hamiltonian in the Jacobi coordinates can be written as:

\begin{multline}
\label{int_pot}
\hat{H} =  -\frac{1}{2\mu_{p}R_p}\partial_{R_p}^2R_p-\frac{1}{2\mu_{e}r}\partial_r^2r + \frac{\hat{L}_p^2}{2\mu_{p}R_p^2}\\ + \frac{\hat{L}_{e}^2}{2\mu_{e}r^2} + \frac{1}{R_p} - \frac{1}{r_1} -\frac{1}{r_2}
\end{multline}
Here, $R_p$ is the distance between the two protons, and $r$ is the distance between the electron and the proton-proton centre of mass.\\
The Hamiltonian is written in atomic units, and the reduced masses are given by
$\mu_p = \frac{m_p}{2}$ and 
$\mu_e =\frac{2m_p}{2m_p+1}$\\
After rescaling $rR_p\Psi = \psi$, the K.E. operator acting on $\psi$ is:
\begin{equation}
\hat{T} =  -\frac{1}{2\mu_p}\frac{\partial^2}{\partial R_p^2}-\frac{1}{2\mu_e}\frac{\partial^2}{\partial r^2} + \frac{\hat{L}_p^2}{2\mu_{p}R_p^2} + \frac{\hat{L}_{e}^2}{2\mu_{e}r^2} 
\end{equation}
The hyperradius $R$ is defined here through the equation $\mu_{pH}R^2=\mu_er^2+\mu_pR_p^2$, and $\mu_{pH}=\frac{m_pm_H}{m_p+m_H}$. As we are only interested in the $L=0$ subspace, one can further substitute $\hat{L_{p}}^2=\hat{L_{e}}^2\equiv\hat{L}^2$, and the two hyperangles are defined as $\cos{\theta} = \hat{R}_p.\hat{r}$ and $\tan{\alpha}=\sqrt{\frac{\mu_e}{\mu_p}}\frac{r}{R_p}$. Finally, the Hamiltonian in hyperspherical coordinates is written as:\\
\begin{multline}
\hat{H} = -\frac{1}{2\mu_{pH}}\left(\frac{1}{R}\frac{\partial}{\partial R}R\frac{\partial}{\partial R}+\frac{1}{R^2}\frac{\partial^2}{\partial \alpha^2}\right) +\\ \frac{\hat{L}^2(\theta)}{2\mu_{pH}R^2}\left(\frac{1}{\sin^2{\alpha}}+\frac{1}{\cos^2{\alpha}}\right) + V(R,\theta,\alpha)
\end{multline}
We separate the hyperradial kinetic energy, and the rest of the terms are combined to form the adiabatic Hamiltonian, which is diagonalized at fixed values of R to obtain the adiabatic potential curves. 
\begin{align}
H_{ad}(R,\Omega)\phi_{\nu}(R;\Omega)=U_{\nu}(R)\phi_{\nu}(R;\Omega)
\end{align}
where $\Omega$ is the set of hyperangles, $\nu$ is the channel index to label the eigenstates $\phi_{\nu}(R,\Omega)$ and $U_{\nu}(R)$ are the fixed R eigenvalues. 
Treating the hyperradius as an adiabatic parameter, we expand our solution in a complete basis
\begin{align}
\psi_E(R,\Omega) = \sum_\nu \frac{F_{E,\nu}(R)}{\sqrt{R}}\phi_{\nu}(R;\Omega)
\end{align}
The $R^{-1/2}$ factor eliminates the first derivative term in the hyperradial kinetic energy operator but also leads to an extra $R^{-2}$ term, which is added to the adiabatic Hamiltonian. The expansion coefficients $F_{E,\nu}(R)$ are found by solving the coupled hyperradial Schr\"{o}dinger equations $\langle{\phi_{\nu}}|\hat{H}-E|{\psi_E}\rangle_{\Omega}=0$, leading to:
\begin{multline}
\label{eq:coupleequation}
\left(\frac{-1}{2\mu_{pH}}\frac{\partial^2}{\partial R^2}+U_{\nu}(R)-E\right)F_{E,\nu}(R) \\-\frac{1}{2\mu_{pH}}\sum_{\nu'}\left(2P_{\nu\nu'}(R)\frac{\partial}{\partial R} + Q_{\nu\nu'}(R)\right) F_{E,\nu'}(R)=0
\end{multline}
where $P_{\nu\nu'}=\langle{\phi_{\nu}}|{\frac{\partial \phi_{\nu'}}{\partial R}}\rangle_{\Omega}$ and $Q_{\nu\nu'}=\langle{\phi_{\nu}}|{\frac{\partial^2 \phi_{\nu'}}{\partial R^2}}\rangle_{\Omega}$ represent the non-adiabatic couplings and $\langle{f}|{g}\rangle_{\Omega}$ represents integrating over only the hyperangular degrees of freedom.\\
The adiabatic Hamiltonian is written as\\
\begin{multline}
\hat{H}_{ad}= -\frac{1}{2\mu_{pH}R^2}\frac{\partial^2}{\partial \alpha^2}+\frac{\hat{L}^2(\theta)}{2\mu_{pH}R^2}\left(\frac{1}{\sin^2{\alpha}}+\frac{1}{\cos^2{\alpha}}\right)\\ + V(R,\Omega)-\frac{1}{8\mu_{pH}R^2}
\end{multline}
$V(R,\Omega)$ is the sum of pairwise Coulomb interactions and has a singularity at $(\theta=0,\alpha=\alpha_c)$, where $\tan{\alpha_c}=\sqrt{\frac{\mu_e}{4\mu_p}}$ and obtaining converged potential curves requires careful representation of the derivative discontinuity in the hydrogenic wave function. Hence, another coordinate transformation implemented is$(\alpha\rightarrow\alpha_c+x^3$ and $\cos\theta\rightarrow 1-y^4$) and the volume element in the hyperangular coordinates becomes $ d\Omega = 12y^3x^2 dy \, dx$
\[ \hat{H}_{ad} = \hat{X} + \hat{Y} + V(R,x,y)\]
\begin{align}
\label{eq:Koperatorx}
  \hat{X} = -\frac{1}{2\mu_{pH}R^2}\left(\frac{1}{3x^2}\frac{\partial}{\partial x}\frac{1}{3x^2}\frac{\partial}{\partial x}+\frac{1}{4}\right)
\end{align}
\begin{multline}
  \hat{Y} = -\frac{1}{2\mu_{pH}R^2}\left[\frac{1}{16y^3}\left(\frac{1}{\sin^2(\alpha_c+x^3)} + \frac{1}{\cos^2(\alpha_c+x^3)}\right)\right.\\\left.\times 
  \frac{\partial}{\partial y}(2y-y^5)\frac{\partial}{\partial y}\right] 
\end{multline}
We diagonalize the adiabatic Hamiltonian $H_{ad}(R,x,y)$ in the basis of B-splines with 140 basis functions in each of the two coordinates. The convergence of the potential curves with respect to the number of basis functions is discussed in the Results section. It is also important to note that one further modification is needed while calculating matrix elements of $\hat{X}$ in any polynomial basis. Consider a matrix element with two B-splines that span the singularity at $x=0$ for the differential operator in $\hat{X}$,  
\[
 \int_{x_{min}}^{x_{max}} b_i(x)\frac{\partial}{\partial x}\left( \frac{1}{3x^2}\frac{\partial b_j(x)}{\partial x}\right) \,dx\
\]
\[
= -\int_{x_{min}}^{x_{max}} \frac{1}{3x^2}\frac{\partial b_i(x)}{\partial x}\frac{\partial b_j(x)}{\partial x} \,dx\ + \left(\frac{b_i(x)}{3x^2}\frac{\partial b_j(x)}{\partial x}\right)_{x_{min}}^{x_{max}}
\]
The second term evaluated at the boundary is zero since the basis doesn't include any B-splines which are non-zero at the boundaries in $x$.
Observe that any B-splines that go to zero linearly in $x$ would cause divergences of this integral, and should not appear in the wavefunction expansion. To address this, a linear transformation is carried out over the set of non-zero B-splines at $x=0$ to eliminate the linear terms in $x$. Say, $\{y_i\}$ is the set of splines that have a non-zero derivative at $x=0$. Define a linear transformation over $\{y_i\}$ to obtain a new set of basis functions $\{u_i\}$ such that $u_i = \sum_k a_{ik}y_k$ and $u_i'(0) = 0$. 
\subsection{\label{subsec:level1} Conversion factors}
Next, consider some factors of $\mu_{ep}$ that will show up in the constants because we do not assume an infinitely heavy nucleus. The ground state energy of the hydrogen atom is given by $E_h=-\frac{m_ee^4}{2(4\pi\epsilon)^2\hbar^2}$, which is -0.5 in atomic units. Replacing $m_e$ with $\mu_{ep}$ we get $E_h=-0.5\mu_{ep}=-0.4997278397$ in atomic units.
Similarly the Bohr-radius $a_0\propto\ 1/m_e\rightarrow a_0/\mu_e$ and the static polarizability of the $1s$ state of hydrogen $\alpha = \frac{9a_0^3}{2} \rightarrow \frac{\alpha}{\mu_{ep}^3}$. Hence, the ground state is slightly more polarizable than a hydrogen atom with an infinitely massive proton would be.

\subsection{\label{subsec:level1} Adiabatic potential curves}
\begin{figure}[h!]
\begin{frame}{}
  \includegraphics[width=1.0\linewidth]{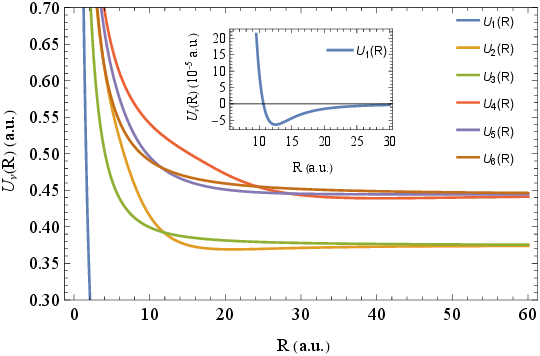} 
  \end{frame}
\caption{\label{fig:pot} The lowest six potential curves for L=0 in the ungerade symmetry as a function of hyperradius $R$. The zero of the energy scale is set to the energy of the hydrogen atom $E_h = -0.4997278397$ in atomic units. Notice that two curves will approach the $n=2$ threshold of hydrogen and three the $n=3$ threshold. The inset shows the lowest potential curve in the ungerade symmetry. The shallow well is responsible for the extremely weakly bound state.}
\end{figure}

The potential curves $U_{\nu}(R)$ were calculated by expanding $\phi_{\nu}(R,\Omega)$ in the B-spline basis and diagonalizing the adiabatic Hamiltonian at different R-values. Since $P_{\nu\nu'}(R)$ is an antisymmetric matrix, the adiabatic corrections to $U_{\nu}(R)$ are given by the diagonal terms of the $Q$ matrix, and the adiabatic potential curves are written as $W_{\nu}(R) = U_{\nu}(R)-\frac{Q_{\nu\nu}(R)}{2\mu_{pH}}$. The potential curves are obtained by diagonalizing in a basis with 140 splines in each of the hyperangular coordinates up to a hyperradius of $R=60a_0$. It is worth noting that as $R$ increases, the convergence of the potential curves begins to deteriorate because the hypernangular extent of the wavefunction gets smaller. At $R=60a_0$ we have an uncertainty of the order of $10^{-10}$a.u. in our curves. The lowest six potential curves of the $L=0$ states for the ungerade symmetry are shown in Fig. \ref{fig:pot}. Asymptotically, they correspond to a hydrogen atom and a far-apart proton and the lowest potential curve $U_1(R)$ can be seen to approach the $1s$ threshold of hydrogen as R increases. The matrix elements $P_{\mu\nu}(R)$  are sharply peaked near the avoided crossings; a Landau-Zener analysis suggests a higher likelihood for adiabatic transitions to occur. Thus, the potential curves are diabatically connected through the crossings as illustrated in Fig. \ref{fig:pot}.  

\subsection{\label{subsec:level1} Post-Adiabatic Theory}

The post-adiabatic theory\cite{klar1976post,klar1977adiabatic} is utilized here to describe the non-adiabatic effects on the adiabatic potential curves. This treatment transforms away the non-adiabatic couplings to give effective energy-dependent potential curves. An advantage of the post-adiabatic approximation is that we can solve an ordinary differential equation (ODE) in an energy-dependent effective potential instead of solving coupled differential equations. Calculation of the effective potentials reduces to solving the following eigenvalue problem:
\begin{align}
\label{eq:padi_evalue}
\begin{pmatrix}
\mathbf{U} + \frac{3}{2\mu_{pH}}\mathbf{P}^2 & \ -2\mathbf{P}\left(\mathbf{1}E-\mathbf{U} + \frac{1}{2}\mathbf{P^2}\right)\\
\mathbf{P} & \mathbf{U}-\frac{1}{2\mu_{pH}}\mathbf{P}^2
\end{pmatrix}\begin{pmatrix}
\Vec{V}_{\nu}\\
\Vec{W}_{\nu}
\end{pmatrix} = U^{eff}_{\nu}\begin{pmatrix}
\Vec{V}_{\nu}\\
\Vec{W}_{\nu}
\end{pmatrix} 
\end{align}
The eigenvalues $U^{eff}_{\nu}(R)$ are the post-adiabatic channel potentials replacing the adiabatic potentials $U_{\nu}(R)$. Note that Eq.\eqref{eq:padi_evalue} differs from the equation given in Ref.\cite{klar1976post,klar1977adiabatic} by a minus sign on the off-diagonal terms, which might have been a typographical error in those references or a different convention; however, the eigenpotentials are the same in either convention. After $\Vec{W}_{\nu}$ are eliminated and  the terms in $\mathbf{P}$ are treated perturbatively,
if the couplings are small, i.e. $\frac{1}{2\mu_{pH}}|P_{\nu\nu'}(R)|^2<<|U_{\nu}(R)-U_{\nu'}(R)|$, this gives: 
\begin{align}
\label{eq:padi_pert}
U_{\nu}^{eff}(R,E) = W_{\nu}(R) + \frac{2}{\mu_{pH}}(E-U_{\nu}(R))\sum_{\mu}\frac{P_{\nu\mu}^2(R)}{U_{\nu}(R)-U_{\mu}(R)}
\end{align}
The $\mu = \nu$ term in the summation doesn't contribute because $P_{\nu\nu} = 0$. The truncated effective potentials are designated as $U_1^{eff}(R,E=0)^{(n)}$,
\begin{align}
\label{eq:postadi}
 U_{\nu}^{eff}(R,0)^{(n)}=W_{\nu}(R) -\frac{2U_\nu}{\mu_{pH}}\sum_{\mu}^n \frac{P_{\nu\mu}^2(R)}{U_{\nu}(R)-U_{\mu}(R)} ,
\end{align} 

which can be analyzed to study the non-adiabatic effects. Since our interest is in near-threshold physics, all the calculations with the post-adiabatic potentials have been done at $E=0$ for $U_1^{eff}(R,E)^{(n)}$.

\subsection{\label{subsec:level1} Streamlined R-matrix solution of the hyperradial equation}
Using the eigenchannel R-matrix method, the negative log derivative $b\equiv -\dfrac{1}{\psi}\dfrac{\partial \psi}{\partial R}$ can be obtained variationally by solving a generalized eigenvalue problem.\cite{aymar1996multichannel}. The problem is set up in a B-spline basis in the hyperradial coordinates.
\begin{align}
\mathbf{\Gamma}\Vec{C} = \overline{b}\mathbf{\Lambda}\Vec{C}
\end{align}
Here $\overline{b}$ is the mass-scaled negative log derivative defined as $\overline{b}=b/\mu_{pH}$. The matrix $\Lambda$ is an overlap surface integral and $\Gamma$ can be written in terms of the overlap matrix $S$, the Hamiltonian $H$ and the Bloch operator $L$.  $H$ is the Hamiltonian operator which acts on the rescaled wave function $\sqrt{R}\psi=R^{5/2}\sin{\alpha}\cos{\alpha}\Psi$ such that:
\begin{subequations}
\begin{multline}
H^{\mu\nu} = \delta_{\mu\nu}\left( -\dfrac{1}{2\mu_{pH}}\dfrac{\partial^2}{\partial R^2} + U_{\nu}(R)\right)\\-\dfrac{1}{2\mu_{pH}}\left(2P_{\mu\nu}(R)\dfrac{\partial}{\partial R} + Q_{\mu\nu}(R)\right)
\end{multline}
\begin{equation}
L^{\mu\nu}_{ij}=\delta_{\mu\nu}\frac{1}{2\mu_{pH}}y^{\mu}_i(R_o)\dfrac{\partial y^{\nu}_j}{\partial R}(R_o) 
\end{equation}
\begin{equation}
\Gamma_{ij}^{\mu\nu}=2(ES_{ij}^{\mu\nu}\delta_{\mu\nu}-H_{ij}^{\mu\nu}-L_{ij}^{\mu\nu})\end{equation}
 \begin{equation}\Lambda_{ij}^{\mu\nu} = \delta_{\mu\nu} y_i^{\mu}(R_o)y_j^{\nu}(R_o) \end{equation} 
 \end{subequations}
 Here $y_i^{\mu}$ is a B-spline in the $\mu$ channel and $R_o$ is the R-matrix boundary.
We use the streamlined R-matrix method \cite{greene1988streamlined} to calculate the log derivative analytically as a function of energy at the boundary of the R-matrix volume. For scattering length calculations, the lowest channel $W_1(R)$ is open; this corresponds to including a spline in our basis set, which is non-zero at the R-matrix boundary. In the streamlined picture, the equation determining the eigenvalues $b$ is given by:
\begin{align}
\mathbf{\Omega}\Vec{C}_O=\overline{b}\mathbf{\Lambda}_{OO}\Vec{C}_O
\end{align}
where $ \mathbf{\Omega}= \mathbf{\Gamma}_{OO}-\mathbf{\Gamma}_{OC}\mathbf{\Gamma}_{CC}^{-1}\mathbf{\Gamma}_{CO}$.
Therefore, in the case of only one open channel, we can write $b$ as a function of energy:
\begin{align}
\frac{b(E)}{\mu_{pH}} = \Gamma_{OO}^{kk}-\frac{1}{2}\sum_{i,j,\lambda}\frac{\Gamma_{OC}^{ki}\chi_{i\lambda}\chi^{\dag}_{\lambda j}\Gamma_{CO}^{jk}}{E-E_{\lambda}}
\end{align}
Here, $i,j$  are compound indices for the basis functions, and $k$ is a compound index for the non-zero spline at the boundary. $E_{\lambda}$ and $\chi_{ij}$ are eigenvalues and eigenvectors of the generalized eigenvalue problem:
\[\sum_j H_{CC}^{ij}\chi_{j\lambda} = E_{\lambda}\sum_j S_{CC}^{ij}\chi_{j\lambda}\]
Next, the procedure for matching the logarithmic derivative at an intermediate R-value is described.  
The hyperradial solutions in the lowest potential curve satisfy the following Schr\"{o}dinger equation at zero energy outside the R-matrix volume:
\begin{align}
\left(-\frac{1}{2\mu_{pH}}\frac{\partial^2}{\partial R^2} -\frac{C}{R^4}\right)f(R) = 0
\end{align}
here $C=\frac{\alpha}{2\mu_{ep}^3}$, $f_0(R) = R\sin{\left(\frac{\sqrt{2C\mu_{pH}}}{R}\right)}$ and $g_0(R)=R\cos{\left(\frac{\sqrt{2C\mu_{pH}}}{R}\right)}$ are two linearly independent solutions. We analytically construct the linear combination of $f_0$ and $g_0$ at zero energy such that the solution has the correct asymptotic form at large R. This procedure allows us to diagonalize $H_{CC}$ in a relatively smaller R-matrix volume while still having the desired asymptotic form at $R\rightarrow\infty$. Say,
\[u(R) = f_0(R) - g_0(R)K\]
Now $\lim_{R\rightarrow\infty}u(R)\propto 1-\frac{R_{pH}}{a}=1-\sqrt{\frac{\mu_{pH}}{\mu_p+\frac{\mu_e}{4}}}\frac{R}{a}= 1-\frac{R}{a}$, where a is the scattering length and $R_{pH}$ is the distance of the proton from the proton-electron centre of mass. This gives us:
\begin{align}
\label{eq:asc}
a=\frac{\sqrt{2\mu_{pH}C}}{K}
\end{align}
And matching $u(R)$ with the variationally calculated $b(E=0)$ at the R-matrix boundary, $R_0$:
\begin{align}
\label{eq:tand}
K=\frac{f_0(R_0)b(0)+f_0'(R_0)}{g_0(R_0)b(0)+g_0'(R_0)}
\end{align}
Using \eqref{eq:asc} and \eqref{eq:tand}, we can finally write the scattering length $a$ as:
\begin{align}
a = \sqrt{2C\mu_{pH}}\left(\frac{g_0(R_0)b(0)+g_0'(R_0)}{f_0(R_0)b(0)+f_0'(R_0)}\right)
\end{align}

The streamlined R-matrix method is also implemented to calculate the vibrational bound state energies. We can match $b(E)$ at $R_0$ to the log derivative of the solution that exponentially decays and is negligible beyond some suitably large R, which can be easily estimated using the WKB wavefunction. 

\section{\label{sec:level3}Results and discussion}
\subsection{\label{subsec:convergence} Convergence of the adiabatic potentials}
\begin{figure}%
    \centering
    \subfigure[]{{\includegraphics[width=0.9\linewidth]{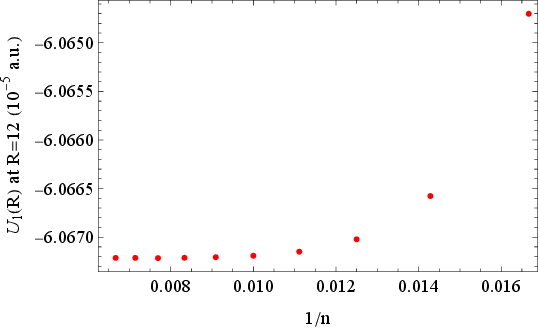} }}%
    
    \subfigure[]{{\includegraphics[width=0.9\linewidth]{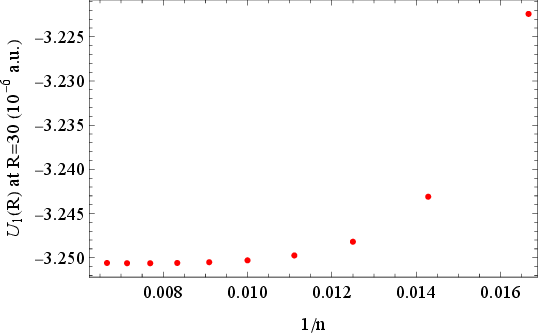} }}%
    \caption{The figures show the lowest potential curve $U_1(R)$ plotted against $1/n$ at two different R values. $n$ is the number of basis functions included in each dimension; the plots contain the same number of basis functions in each dimension. Figures (a) and (b) are for $U_1(R)$ at $R=12$ a.u. and $R=30$ a.u. respectively.}%
    \label{fig:convpot}%
\end{figure}

The adiabatic Hamiltonian is diagonalized in the basis of B-splines. In practice, a basis size of 140 functions in each dimension works well to obtain converged potentials, with an error of the order of $10^{-12}$ atomic units. The accuracy of the eigenvalue solver limits the ability to achieve convergence to more decimal places, and it appears that higher accuracy would require an implementation of the entire calculation in quadruple precision. Fig.\ref{fig:convpot} shows the lowest potential curve $U_1(R)$ versus the number of basis functions at two different R-values, close to the minima of the potential and at a relatively larger hyperradius to show convergence. 
\subsection{\label{subsec:boundencalc} Bound state calculations}
The lowest adiabatic potential curve $W_{1}(R)=U_{1}(R) - \frac{Q_{11}(R)}{2\mu_p}$ obtained numerically up to $R=60a_0$, is smoothly joined to the expected long-range behaviour of the polarizability potential $-\frac{\alpha}{\mu_{ep}^3R^4}$ plus the exact dissociation threshold, as has been shown by Macek\cite{J_Macek_PQasymptotics}. Here $\alpha$ is the static polarizability of the $1s$ state of hydrogen and $\mu_{ep} = \frac{m_p}{1+m_p}$. The hyperradial Schr\"{o}dinger equation for the single and multichannel case is solved by diagonalizing Eq. \eqref{eq:coupleequation} in the basis of B-splines with vanishing boundary conditions at $R_{min}=0$ and $R_{max}=10000a_0$. In the three-channel case, the non-adiabatic couplings between the second and third channels ($P_{23}$ and $Q_{23}$) are neglected, as they would only contribute to the higher orders. The energies are given in Table \ref{coup-table}. The ground state energy for $W_1(R)$, which only has the diagonal adiabatic correction $Q_{11}(R)$ added to it, comes out to $E_0=-1.56622\times10^{-5}$ a.u., and this also gives an upper-bound on the energy of this state as shown in Ref. \cite{upper_lowerbound}. This is already close to the actual value of $E_0 = -1.56625\times10^{-5}$ a.u. from Ref. \cite{hilico2001polarizabilities}. 
\begin{table}[h!]
\begin{ruledtabular}
\centering
 \begin{tabular}{c c c} 
 Number of channels & $E_0\times10^5$ (a.u.) & $E_1\times10^9$ (a.u.)\\ [0.5ex] 
 \hline
1 & -1.56622 & -1.0734 \\
2 & -1.56623 & -1.0736\\
3 & -1.56623 & -1.0738 \\ [1ex] 
 \end{tabular}
 \end{ruledtabular}
 \caption{Ground and excited state energies in $W_1(R)$ from the coupled channel equation}
 \label{coup-table}
\end{table}

\begin{figure}[h!]
\begin{frame}{}
  \includegraphics[width=1.0\linewidth]{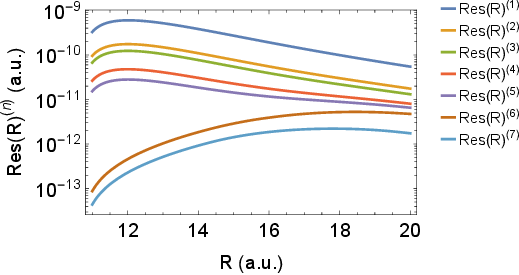} 
  \end{frame}
\caption{\label{fig:padires} This shows a log plot of $Res(R)^{(n)}$. $Res(R)^{(n)}=U^{eff}_1(R,0)^{(n)}-U^{eff}_1(R,0)^{(8)}$, which clearly shows the convergence of $U^{eff}_1(R)^{(n)}$ as n increases.} 
\end{figure}

Next, consider the truncated effective potentials defined in \eqref{eq:postadi} from the post-adiabatic approximation and diagonalize $U^{eff}_1(R,0)^{(n)}$ up to $n=8$ in the B-spline basis. Fig. \ref{fig:padires}, plots $Res(R)^{(n)}=U^{eff}_1(R,0)^{(n)}-U^{eff}_1(R,0)^{(8)}$ as $n$ increases, showing how $U^{eff}_1(R,0)^{(n)}$ starts converging. Table \ref{padres-table} summarizes the energies for the ground and the excited state. Although the energy of the first excited state varies slowly with $n$, it only differs from $E_1=-1.85045\times10^{-9}$ a.u.\cite{carbonell2003new} by about a percent, and the energy of the ground state matches the value from Ref. \cite{hilico2001polarizabilities} to many decimal places. It should be noted that $U^{eff}_1(R,0)^{(0)} = W_1(R)$, so the first row of Table \ref{padres-table} is the same as that of Table \ref{coup-table}.

\begin{figure}[h!]
\begin{frame}{}
  \includegraphics[width=1.0\linewidth]{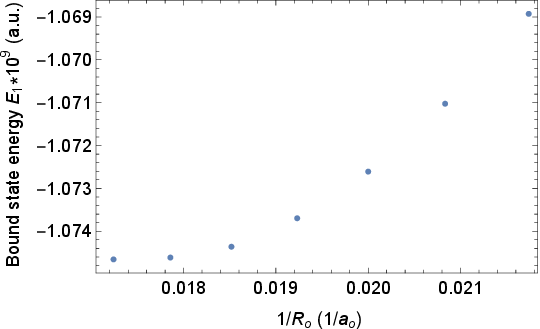} 
  \end{frame}
\caption{\label{fig:exben} The figure shows the variation of the bound state energy of the first excited state with the inverse box size of the R-matrix $1/R_o$. The bound state energy starts converging as the box size $R_o$ increases.} 
\end{figure}

\begin{table}[h!]
\begin{ruledtabular}
\centering
 \begin{tabular}{c c c} 
 No. of terms & $E_0\times10^5$ (a.u.) & $E_1\times10^9$ (a.u.)\\ [0.5ex] 
 \hline
0 & -1.56622 & -1.0734 \\
1 & -1.56623 & -1.0736 \\
2 & -1.56625 & -1.0738 \\
3 & -1.56625 & -1.0738 \\
4 & -1.56625 & -1.0739 \\
5 & -1.56625 & -1.0739 \\
6 & -1.56625 & -1.0739 \\
7 & -1.56625 & -1.0739 \\
8 & -1.56625 & -1.0739 \\  
 \end{tabular}
 \end{ruledtabular}
 \caption{Ground and excited state energies (in atomic units) obtained from the post-adiabatic approximation}
 \label{padres-table}
\end{table}
For bound state calculations, the semi-analytical form of the logarithmic derivative $b_(E)$ from the streamlined R-matrix method can also be utilized and matched with the exponentially decaying solution that is propagated inwards in the pure polarizability potential from large distances. The energy at which the two log derivatives match is the bound state eigenenergy. Finally, the R-matrix box size $R_o$ is varied to test for convergence with respect to that key parameter of our calculation. The energy of the first excited state $E_1$ for the case of three channels is plotted in Fig. \ref{fig:exben}. This gives us an energy $E_1=-1.0745\times10^{-9}$ a.u., which differs slightly from the energies we get from the other two methods. We suspect this is because the adiabatic potential curves are slightly less accurately converged for larger hyperradius.

\subsection{\label{subsec:scatlencalc} Scattering length calculations}
\begin{figure}[h!]
\begin{frame}{}
  \includegraphics[width=1.0\linewidth]{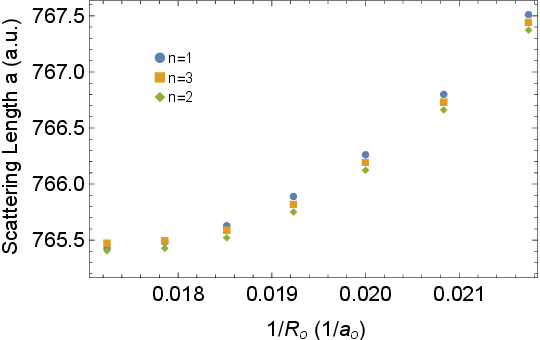} 
  \end{frame}
\caption{\label{fig:scatteringlen} The variation of the scattering length is shown as a function of the inverse of the box radius, $1/R_0$, for the H$^+$ + H(1s) collision with 1, 2 or 3 channels. It can be seen that the scattering length starts converging as the box size $R_0$ is increased.}
\end{figure}

In keeping with the streamlined eigenchannel R-matrix method the closed-closed partition of the Hamiltonian is constructed and diagonalized using the previously mentioned basis, including up to three adiabatic (diabatized) hyperspherical channels. Again, the non-adiabatic couplings between the second and third channels ($P_{23}(R)$ and $Q_{23}(R)$) are neglected for the three-channel calculation. After extracting the log derivative, the scattering length is extracted using the previously outlined matching procedure. The R-matrix boundary, $R_0$ is varied to test the convergence, and the results are depicted in Fig. \ref{fig:scatteringlen}
\begin{table}[h!]
\centering
\begin{ruledtabular}
 \begin{tabular}{c c} 
  & Scattering length $a$ (a.u.) \\ 
 \hline
 Carbonell \textit{et al.}\cite{lazauskas2002scattering,carbonell2003new} & 750(5) \\
 Bodo \textit{et al.}\cite{bodo2008ultra} & 725.2 \\
 Kaiser \textit{et al.}\cite{kaiser2013quantization} & 762.8 \\
 Zhou \textit{et al.}\cite{zhou2015hyperspherical} & 715.0 \\
 Beyer and Merkt\cite{beyer2018hyperfine} & 757(7)\\
 This work & 765.5(2) \\  
 \end{tabular}
 \end{ruledtabular}
\caption{Scattering length for the H$^+$ + H(1s) collision. This table only includes the works that don't include hyperfine interaction}
\label{scatlentbl}
\end{table}
\subsection{\label{subsec:adhsph} Adiabatic representations}
We emphasize that the adiabatic hyperspherical representation's ability to describe this problem's main features with just the adiabatic effects. The Born-Oppenheimer potential curves go to the wrong threshold, resulting in the diagonal Born-Oppenheimer correction (DBOC) asymptotically going to a constant to correct for this difference. The hyperspherical adiabatic representation doesn't have problems with the electron translation factors, as explained in Ref. \cite{hyperspherical_ETF,macek1987bypassing}. As a result of this, the diagonal adiabatic correction $Q_{11}(R)$ asymptotically goes to zero as $1/R^2$. Around the minima of the $2p\sigma_u$ potential, the DBOC is roughly $57~\mathrm{cm}^{-1}$ for the BO potential, while for the hyperradial potential curve, the diagonal correction is of the order of $0.5~\mathrm{cm}^{-1}$. The calculation for the bound state energy for this extremely weakly bound state with just the adiabatic correction gives $E_1 = -1.0734\times10^{-9}$ a.u., which is an error of about 1$\%$ from the exact value. The scattering length calculation also yields a value of $a = 765.5(2)$, which is comparable to other authors' results for this system. Several values are summarized in Table \ref{scatlentbl}. More importantly, these calculations demonstrate the usefulness of our chosen adiabatic representation for minimizing the non-adiabatic effects. 
\section{Conclusion}
The extremely weakly bound state of the hydrogen molecular ion in the ungerade symmetry for $L=0$, is explored, along with its huge scattering length in the limit of zero collision energy. The calculations demonstrate the power of the adiabatic hyperspherical representation in describing physics close to the threshold, leading to minimal non-adiabatic effects and an accurate single-potential description. Using just the adiabatic potential, our binding energy for the ungerade excited state differs from the non-adiabatic calculations by only about 1$\%$. Moreover, the post-adiabatic theory is shown in this system to be effective in its ability to incorporate the non-adiabatic effects relevant in this energy range.  Of course, it should be kept in mind that for the description of physical processes at higher energies, where inelastic collisions can transfer probability between the adiabatic channels, the post-adiabatic method becomes less useful.
\label{sec:Conclusion}

\section{Acknowledgment}
This work was supported by the U.S. Department of Energy, Office of Science, Basic Energy Sciences, under Award No. DE-SC0010545.
\section{Article information}
\subsection{Data availability}
Data generated or analyzed during this study are available from the corresponding author upon reasonable request.
\label{sec:art_info}
\section{Author information}
\subsection{Competing interests}
The authors declare there are no competing interests.
\label{sec:auth_info}
\section{Appendix}
\subsection{Perturbative expansion for the post-adiabatic potential}
This section provides a detailed derivation of the perturbative expression for the post-adiabatic potential we used in the paper. We start by rewriting Eq.\eqref{eq:coupleequation} in matrix form:
\begin{align}
\left[\left(\mathbf{1}\frac{\partial}{\partial R} + \mathbf{P}\right)^2 + 2\left(\overline{E}\mathbf{I}-\overline{\mathbf{U}}\right)\right]\Vec{F}=0
\end{align}
where, $\overline{E}=\mu_{pH}E$ is the rescaled energy and $\overline{\mathbf{U}}=\mu_{pH}\mathbf{U}$ is the rescaled potential matrix which is diagonal. Now this equation is transformed into a first-order system by substituting $-\left(\mathbf{1}\frac{\partial}{\partial R} + \mathbf{P}\right)\Vec{F}=\Vec{G}$ to get:
\begin{align}
\begin{pmatrix}
\mathbf{1}\partial_R + \mathbf{P} & \mathbf{1}\\
-2\left(\mathbf{1}\overline{E}-\overline{\mathbf{U}}\right) & \mathbf{1}\partial_R + \mathbf{P}
\end{pmatrix}\begin{pmatrix}
\Vec{F}\\
\Vec{G}
\end{pmatrix} = 0
\end{align}
Now, we want to consider a linear transformation $\mathbf{T}$ which transforms away the couplings:
\begin{align}
\begin{pmatrix}
\Vec{F}\\
\Vec{G}
\end{pmatrix}=\mathbf{T}\begin{pmatrix}
\Vec{f}\\
\Vec{g}
\end{pmatrix}=\begin{pmatrix}
\mathbf{V} & \mathbf{W}\\
\mathbf{X} & \mathbf{Y}
\end{pmatrix}\begin{pmatrix}
\Vec{f}\\
\Vec{g}
\end{pmatrix}
\end{align}
Such that the transformed $\Vec{f}$ and $\Vec{g}$ are uncoupled to within terms of the order $\partial_R \mathbf{T}$ 
\begin{equation}\label{eq:star1}
\partial_R \Vec{f} = -\Vec{g} + \mathcal{O} (\partial_R \mathbf{T}) 
\end{equation}
\begin{equation}\label{eq:star2}
\partial_R \Vec{g} = 2(\mathbf{1}\overline{E}-\overline{\mathbf{u}})\Vec{f} +  \mathcal{O} (\partial_R \mathbf{T})
\end{equation}
where $\overline{\mathbf{U}}^{eff}$ is a diagonal matrix of the transformed post-adiabatic potentials. Substituting this, we get:
\begin{equation}
\begin{pmatrix}
\mathbf{1}\partial_R + \mathbf{P} & \mathbf{1}\\
-2\left(\mathbf{1}\overline{E}-\overline{\mathbf{U}}\right) & \mathbf{1}\partial_R + \mathbf{P}
\end{pmatrix}\begin{pmatrix}
\mathbf{V} & \mathbf{W}\\
\mathbf{X} & \mathbf{Y}
\end{pmatrix}\begin{pmatrix}
\Vec{f}\\
\Vec{g}
\end{pmatrix} = 0
\end{equation}
Neglecting terms of the order of $\partial_R \mathbf{T}$, we can write these equations as:
\begin{align}\label{eq:star3}
\mathbf{V} \partial_R \Vec{f} + \mathbf{PV}\Vec{f} + \mathbf{X}\Vec{f} + \mathbf{W} \partial_R \Vec{g} + \mathbf{PW} \Vec{g} + \mathbf{Y} \Vec{g} = 0
\end{align}
\begin{multline}\label{eq:star4}
-2(\mathbf{1}\overline{E}-\overline{\mathbf{U}})\mathbf{V}\Vec{f} + \mathbf{X}\partial_R\Vec{f} + \mathbf{PX}\Vec{f} -2(\mathbf{1}\overline{E}-\overline{\mathbf{U}}^{eff})\mathbf{W}\Vec{g}\\ + \mathbf{Y}\partial_R \Vec{g} + \mathbf{PY} \Vec{g} = 0
\end{multline}
Using Eq. \ref{eq:star1} and \ref{eq:star2} in Eq. \ref{eq:star3} we get $\mathbf{X}=-2\mathbf{W}(\mathbf{1}\overline{E}-\overline{\mathbf{U}}^{eff}) - \mathbf{PV}$ and $\mathbf{Y}=-\mathbf{PW}+\mathbf{V}$. Eliminating $\mathbf{X}$ and $\mathbf{Y}$, we get equations for each column of $\mathbf{V}$ and $\mathbf{W}$ satisfying
\begin{align}
\label{eq:padibef}
\begin{pmatrix}
\overline{\mathbf{U}}-\mathbf{1}\overline{U}^{eff}_{\nu}-\dfrac{1}{2}\mathbf{P}^2 & -2\mathbf{P}\left(\overline{E}-\overline{U}^{eff}_{\nu}\right)\\
\mathbf{P} & \overline{\mathbf{U}}-\mathbf{1}\overline{U}^{eff}_{\nu}-\dfrac{1}{2}\mathbf{P}^2
\end{pmatrix}\begin{pmatrix}
\Vec{V}_{\nu}\\
\Vec{W}_{\nu}
\end{pmatrix} = 0
\end{align}
Multiply from the left by $\begin{pmatrix}
\mathbf{1} & 2\mathbf{P}\\
\mathbf{0} & \mathbf{1}
\end{pmatrix}$ and solving for the post-adiabatic potential $\overline{\mathbf{U}}^{eff}(R)$ reduces to solving the following eigenvalue problem, same as Eq. \ref{eq:padi_evalue}:
\begin{align}
\begin{pmatrix}
\mathbf{U} + \frac{3}{2\mu_{pH}}\mathbf{P}^2 & \ -2\mathbf{P}\left(\mathbf{1}E-\mathbf{U} + \frac{1}{2\mu_{pH}}\mathbf{P^2}\right)\\
\mathbf{P} & \mathbf{U}-\frac{1}{2\mu_{pH}}\mathbf{P}^2
\end{pmatrix}\begin{pmatrix}
\Vec{V}_{\nu}\\
\Vec{W}_{\nu}
\end{pmatrix} = U^{eff}_{\nu}\begin{pmatrix}
\Vec{V}_{\nu}\\
\Vec{W}_{\nu}
\end{pmatrix} 
\end{align}
In order to obtain the perturbative form given in Eq. \ref{eq:padi_pert}, first eliminate $\Vec{W}_{\nu}$ from Eq. \ref{eq:padibef} to obtain
\begin{multline}
\left[\mathbf{U}-\dfrac{1}{2\mu_{pH}}\mathbf{P}^2+2\left(E-U^{eff}_{\nu}\right)\right.\\\left.\times\mathbf{P}\left( \mathbf{U}-\dfrac{1}{2\mu_{pH}}\mathbf{P}^2-\mathbf{1}U^{eff}_{\nu}\right)^{-1}\mathbf{P}\right]\Vec{V}_{\nu} = U^{eff}_{\nu}\Vec{V}_{\nu}
\end{multline}
Treating terms perturbatively in powers of $\mathbf{P}$, in the limit of weak coupling  $\frac{1}{2\mu_{pH}}|P_{\nu\nu'}(R)|^2<<|U_{\nu}(R)-U_{\nu'}(R)|$, we obtain Eq. \ref{eq:padi_pert}:
\begin{align}
U_{\nu}^{eff}(R,E) = W_{\nu}(R) + \frac{2}{\mu_{pH}}(E-U_{\nu}(R))\sum_{\mu}\frac{P_{\nu\mu}^2(R)}{U_{\nu}(R)-U_{\mu}(R)}
\end{align}
where $W_{\nu}(R) = U_{\nu}(R)-P^2_{\nu\nu}(R)/2\mu_{pH}= U_{\nu}(R)-Q_{\nu\nu}(R)/2\mu_{pH}$


\bibliographystyle{unsrt} 
\bibliography{bibliography} 

\end{document}